\documentclass[aps,pre,reprint,10pt,superscriptaddress,showpacs]{revtex4-1}
\usepackage{amsfonts} 
\usepackage{amsmath}
\usepackage{amssymb}
\usepackage{graphicx}%\[  \]
\usepackage[caption=false]{subfig}
\usepackage{color}
\usepackage{enumerate}
\usepackage[]{natbib}
\begin{document}
\let\WriteBookmarks\relax
\def\floatpagepagefraction{1}
\def\textpagefraction{.001}
%%%

%\shorttitle{Stability of vertically stratified fluids}
%\shortauthors{T.D. Kaladze et~al.}
%%%
\title{Thermal expansion of atmosphere and stability of vertically stratified fluids}
%%%%%%%%%%%%
%\ead[url]{www.cvr.cc, cvr@sayahna.org}
%\credit{Conceptualization of this study, Methodology, Software}
\author{T. D. Kaladze}
\email{tamaz_kaladze@yahoo.com}
\affiliation{ I. Vekua Institute of Applied Mathematics and E. Andronikashvili Institute of Physics, Tbilisi State University, Georgia}
\author{A.P. Misra} 
\email{apmisra@visva-bharati.ac.in; apmisra@gmail.com}
\affiliation{Department of Mathematics, Siksha Bhavana, Visva-Bharati University, Santiniketan-731 235, India}

\begin{abstract}
%\begin{linenumbers}
  The influence of thermal expansion of the Earth's atmosphere on the stability of vertical stratification of fluid density and temperature is studied. We show  that such an influence leads to  the instability of incompressible flows. Modified by the thermal expansion coefficient, a new expression for the Brunt-V{\"a}is{\"a}l{\"a} frequency is derived, and a critical value of the thermal expansion coefficient for which the instability occurs is revealed. 
%\end{linenumbers}         
\end{abstract}

%\item Generation of thermoacoustic shocks  in complex plasmas is shown
%\item Evolution of shocks is governed by Burgers equation
%\item Shocks may be damped or anti-damped
%\item Both analytical and numerical shock solutions are presented
%\begin{keywords}
%Thermal expansion \sep Internal gravity wave \sep Instability \sep Atmospheric fluids \sep Brunt-V{\"a}is{\"a}l{\"a} frequency
%\end{keywords}
\maketitle
%\linenumbers
\section{Introduction} \label{sec-intro}
  Climate change is vitally connected to the warming processes (such as convection, in which the heat energy gets transferred by the movement of neutral fluids from one place to another) in the Earth's atmosphere. In addition, numerous other processes, including meteorological and auroral activities and a solar eclipse, can cause equilibrium density and pressure inhomogeneities, and their gradients. As a result, the atmospheric fluids under gravity become stratified, and in the interior, the small-scale density and pressure fluctuations can produce internal gravity waves (IGWs).   The latter are thus of interest in the general circulation of atmospheric stratified fluids \cite{miyoshi2008,plougonven2014}. So, the characteristics of IGWs become the primary investigation of many scientists. Not only do these waves play crucial roles in particle transport and momentum and energy transfers, as they propagate vertically from the Earth's surface to the upper atmosphere, but these are also relevant in large-scale zonal flows \cite{horton2008}, formation of solitary vortices \cite{stenflo1987,misra2021}, and for the emergence of chaos and turbulence \cite{misra2021,shukla2012,mendonca2015}.    In the generation of IGWs, buoyancy plays the role of restoring force that opposes vertical displacements of fluid particles under gravity, and they are associated with the equilibrium density and temperature inhomogeneities.  Typically, the frequency of IGWs ranges in between the Coriolis parameter and the Brunt-V{\"a}is{\"a}l{\"a} frequency, i.e., $10^{-4}~\rm{s^{-1}}<\omega<1.7\times10^{-2}~\rm{s^{-1}}$ and their amplitudes are relatively small in the tropospheric and stratospheric layers \cite{plougonven2014}.  The linear and nonlinear theories of IGWs have been studied by several authors owing to their fundamental importance in understanding the Earth's atmosphere \citep{shukla2012,misra2021,stenflo1987,mendonca2015,kaladze2022,roy2019}.  
\par  
Typically, the dynamics of stratified fluids are more complex than homogeneous fluids. When the stratified fluids are stable, they can support the existence and propagation of various kinds of gravity waves, including IGWs. However, the stratified fluids may become unstable due to the density variations in different layers of the atmosphere. In this situation, the corresponding  Brunt-V{\"a}is{\"a}l{\"a} frequency may become  imaginary due to a negative density gradient, i.e., when the atmospheric fluid density decreases with height \cite{acheson1973}.  In addition, if the temperature variations (spatial) occur due to differential heating and hence the density variations owing to thermal expansion, there may be competitive roles between the temperature and density gradients, and the relevant fluid dynamics becomes more interesting to study.
\par   
   In this letter, we study the influence of thermal expansion on the stability of vertical stratification of atmospheric fluids (in the regions of the troposphere and stratosphere). We show that the Brunt-V{\"a}is{\"a}l{\"a} frequency $N(z)$ gets tightly connected to IGWs, and it stimulates their horizontal propagation.  In the case when $N^2(z)>0$, the background vertical stratification is said to be stable, but when $N^2(z)<0$, the stratification becomes unstable. Also, we discuss the behaviors of $N(z)$ with the effects of the thermal expansion coefficient.       
%%%%%%%%%%%%%
\section{Basic equations and analysis with observational data} \label{sec-basic}
We consider the linear propagation of IGWs in incompressible stratified atmospheric neutral fluids. As a starting point, we consider the following momentum balance and the continuity equations for incompressible neutral fluids.  
  \begin{equation}\label{eq-mom}
 \frac{\partial {\bf u}}{\partial t}+\left({\bf u}\cdot\nabla\right){\bf u}=-\frac{1}{\rho}\nabla p+{\bf g}, 
\end{equation} 
\begin{equation}\label{eq-cont}
\frac{d\rho}{dt}\equiv \frac{\partial \rho}{\partial t}+\left({\bf u}\cdot\nabla\right)\rho=0,\rm{i.e.,}~ \nabla\cdot{\bf u}=0,
\end{equation} 
where ${\bf u}$, $\rho$, and $p$  are the neutral fluid velocity, mass density, and the pressure respectively, and ${\bf g}=(0,0,-g)$ is the constant gravitational acceleration directed vertically downward. In equilibrium without the fluid flow, we have from Eq. \eqref{eq-mom}
\begin{equation}
\frac{\partial p_0}{\partial z}=-\rho_0 g.
\end{equation}

\par
\begin{figure*}
\centering
\includegraphics[width=6.5in, height=3in]{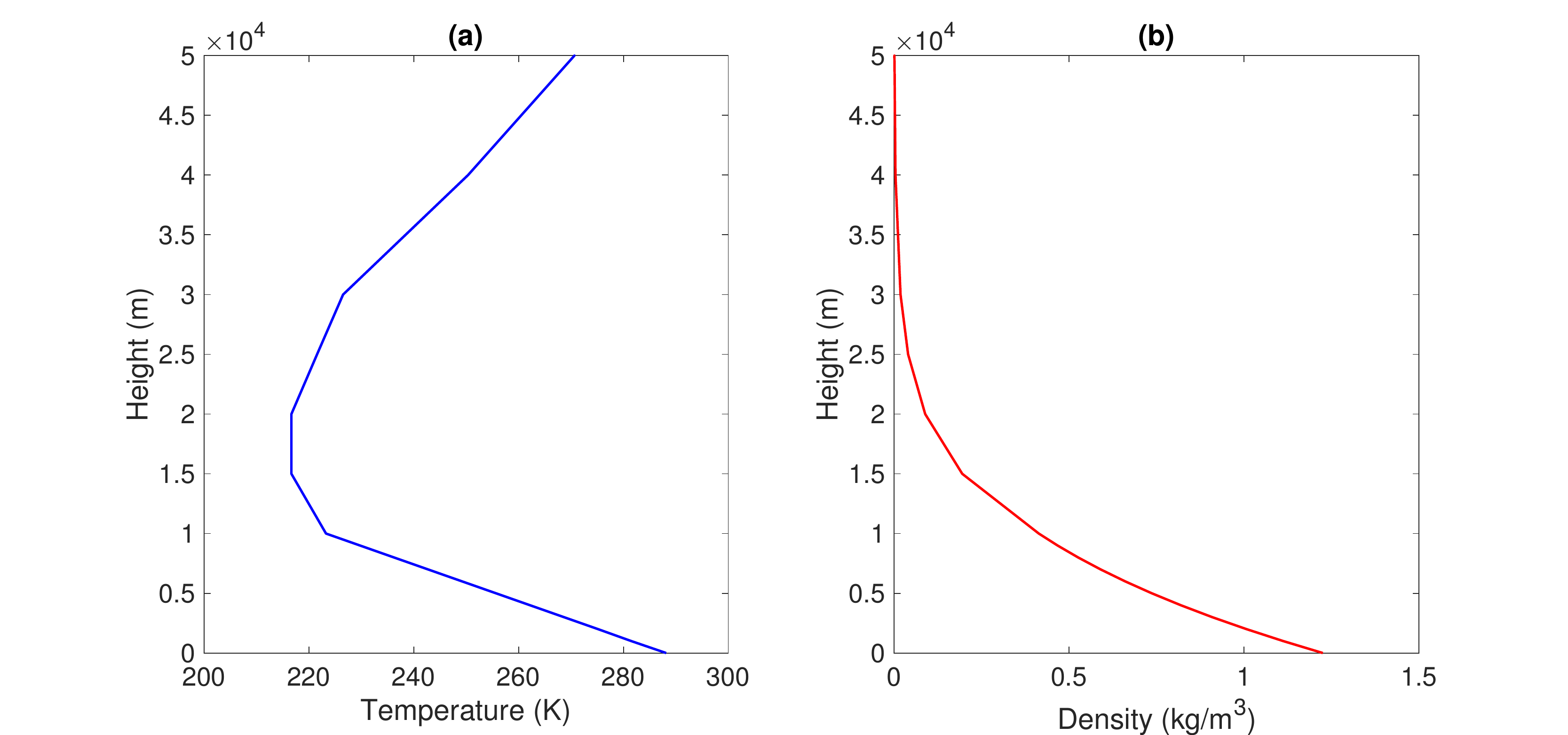}
\caption{The variations of the temperature [subplot (a)] and the density [subplot (b)] with the atmospheric height are shown based on the data as in Table \ref{table1}. }
\label{fig1}
\end{figure*}

 
As said, differential heating causes spatial variations of temperature in the fluid, which in turn  produces the density variation due to the thermal expansion.   Thus, if $\beta~\rm{(K^{-1}})$ is the volumetric thermal expansion coefficient of the heated incompressible fluid,   the equation of state can be written as \cite{acheson1973}
\begin{equation}\label{eq-rho}
\rho=\rho_0(z)\left(1-\beta T\right),
\end{equation}
where $\rho_0$ is the fluid mass density at temperature $T=0$. Considering the data for the ``U.S. Standard Atmosphere Air Properties" \cite{usdata1}, the density and temperature variations of the atmosphere with the height (stratification) are presented in Table \ref{table1} and the variations are graphically exhibited in Fig. \ref{fig1}. In Table \ref{table1}, the temperature and density gradients are obtained using the central difference formula.  From Table \ref{table1} and Fig. \ref{fig1}, it is evident that the fluid density decreases with the height, i.e.,  $d\rho_0/dz<0$ in the whole region of $0<z<50$ (km). However, the temperature decreases with the height, i.e., $dT_0/dz<0$ in the interval $0<z<15$ (km), but the same increases in the other interval, i.e.,  $dT_0/dz>0$ in  $15<z<50$ (km).   So, we are interested mainly in the altitudes of the troposphere (ranging from $0$ to $15$ km) and stratosphere (ranging from $15$ to $50$ km) and consider the vertical distribution of Brunt-V{\"a}is{\"a}l{\"a}  in the neutral fluid atmosphere. 
%%%%%%%%%%%%%%%%%% 
%%%%%%%%%%%%%%%%%%%%%%%
\par 
%Having obtained the temperature and density variations as in  Table \ref{table1} and their graphical representations in Fig. \ref{fig1}, one can   introduce four artificial interfaces: (i) Bottom at $z=0$ (Earth's surface), (ii) Tropopause level $z_1$ at the height $\approx10$ km, (iii) Level $z_3$ at the height of $20$ km of stratosphere, and (iv) Stratopause level $z_3$ at the height of $\approx50$ km between the stratosphere and mesosphere. In the tropospheric interval of $10$ km $(0<z<z_1)$, the temperature $T_0$ decreases with the height from $288$ K till $216$ K, so that $dT_0/dz<0$. The interval between $z_1$ and $z_2$ $(z_1<z<z_2)$ is something special because, $dT_0/dz\approx0$ therein. In the stratospheric interval of $30$ km $(z_2<z<z_3)$, the temperature increases with the height from $216$ K till $270$ K, so that $dT_0/dz>0$.   
%%%%%%%%%%%%%%%%%%%%%%%%%%
\begin{table*} 
\centering
\begin{tabular}{|c c c c c c|} 
\hline
Height\hspace{0.2cm} & Temperature  \hspace{0.2cm} &  Density  \hspace{0.2cm} &  Gravitational  \hspace{0.2cm} &  Temperature gradient  \hspace{0.2cm} & Density gradient    \\
$z$ (m) \hspace{0.2cm} & $T_0$ (K) \hspace{0.2cm} & $\rho_0$ (kg/m$^3$)  \hspace{0.2cm} &   acceleration,  $g$ (m/s$^2$)  \hspace{0.2cm} &   $\left(dT_0/dz\right)$ $\left(\rm{K/m}\right)$  \hspace{0.2cm} &   $\left(d\rho_0/dz\right)$ $\left(\rm{kg/m^4}\right)$\\
$\left(\times10^{3}\right)$ \hspace{0.2cm} & \hspace{0.2cm} & \hspace{0.2cm} & \hspace{0.2cm} & $\left(\times10^{-3}\right)$ \hspace{0.2cm} & $\left(\times10^{-3}\right)$\\
\hline
 $0$ \hspace{0.3cm} & $288.15$ \hspace{0.3cm} &  $1.225$  \hspace{0.3cm} &  $9.807$  \hspace{0.3cm} &  $-1.051$ \hspace{0.3cm} & $-0.0141$ \hspace{0.3cm}  \\
$1$ \hspace{0.3cm} & $281.65$ \hspace{0.3cm} &  $1.112$  \hspace{0.3cm} &  $9.804$  \hspace{0.3cm} &  $-6.5$ \hspace{0.3cm} & $-0.109$ \hspace{0.3cm}  \\
$2$ \hspace{0.3cm} & $275.15$ \hspace{0.3cm} &  $1.007$  \hspace{0.3cm} &  $9.801$  \hspace{0.3cm} &  $-6.495$ \hspace{0.3cm} & $-0.1014$ \hspace{0.3cm}  \\
$3$ \hspace{0.3cm} & $268.66$ \hspace{0.3cm} &  $0.9093$  \hspace{0.3cm} &  $9.797$  \hspace{0.3cm} &  $-6.49$ \hspace{0.3cm} & $-0.0938$ \hspace{0.3cm}  \\
$4$ \hspace{0.3cm} & $262.17$ \hspace{0.3cm} &  $0.8194$  \hspace{0.3cm} &  $9.794$  \hspace{0.3cm} &  $-6.49$ \hspace{0.3cm} & $-0.0864$ \hspace{0.3cm}  \\
$5$ \hspace{0.3cm} & $255.68$ \hspace{0.3cm} &  $0.7364$  \hspace{0.3cm} &  $9.791$  \hspace{0.3cm} &  $-6.49$ \hspace{0.3cm} & $-0.0796$ \hspace{0.3cm}  \\
$6$ \hspace{0.3cm} & $249.19$ \hspace{0.3cm} &  $0.6601$  \hspace{0.3cm} &  $9.788$  \hspace{0.3cm} &  $-6.49$ \hspace{0.3cm} & $-0.0732$ \hspace{0.3cm}  \\ 
$7$ \hspace{0.3cm} & $242.70$ \hspace{0.3cm} &  $0.5900$  \hspace{0.3cm} &  $9.785$  \hspace{0.3cm} &  $-6.49$ \hspace{0.3cm} & $-0.0671$ \hspace{0.3cm}  \\ 
$8$ \hspace{0.3cm} & $236.21$ \hspace{0.3cm} &  $0.5258$  \hspace{0.3cm} &  $9.782$  \hspace{0.3cm} &  $-6.485$ \hspace{0.3cm} & $-0.0614$ \hspace{0.3cm}  \\ 
$9$ \hspace{0.3cm} & $229.73$ \hspace{0.3cm} &  $0.4671$  \hspace{0.3cm} &  $9.779$  \hspace{0.3cm} &  $-6.48$ \hspace{0.3cm} & $-0.0562$ \hspace{0.3cm}  \\ 
$10$ \hspace{0.3cm} & $223.25$ \hspace{0.3cm} &  $0.4135$  \hspace{0.3cm} &  $9.776$  \hspace{0.3cm} &  $-2.18$ \hspace{0.3cm} & $-0.0454$ \hspace{0.3cm}  \\ 
$15$ \hspace{0.3cm} & $216.65$ \hspace{0.3cm} &  $0.1948$  \hspace{0.3cm} &  $9.761$  \hspace{0.3cm} &  $-0.66$ \hspace{0.3cm} & $-0.0325$ \hspace{0.3cm}  \\ 
$20$ \hspace{0.3cm} & $216.65$ \hspace{0.3cm} &  $0.08891$  \hspace{0.3cm} &  $9.745$  \hspace{0.3cm} &  $0.49$ \hspace{0.3cm} & $-0.0155$ \hspace{0.3cm}  \\ 
$25$ \hspace{0.3cm} & $221.55$ \hspace{0.3cm} &  $0.04008$  \hspace{0.3cm} &  $9.730$  \hspace{0.3cm} &  $0.986$ \hspace{0.3cm} & $-0.0071$ \hspace{0.3cm}  \\ 
$30$ \hspace{0.3cm} & $226.51$ \hspace{0.3cm} &  $0.01841$  \hspace{0.3cm} &  $9.715$  \hspace{0.3cm} &  $1.92$ \hspace{0.3cm} & $-0.0024$ \hspace{0.3cm}  \\ 
$40$ \hspace{0.3cm} & $250.35$ \hspace{0.3cm} &  $0.003996$  \hspace{0.3cm} &  $9.684$  \hspace{0.3cm} &  $2.21$ \hspace{0.3cm} & $-0.0009$ \hspace{0.3cm}  \\ 
$50$ \hspace{0.3cm} & $270.65$ \hspace{0.3cm} &  $0.001027$  \hspace{0.3cm} &  $9.654$  \hspace{0.3cm} &  $0.945$ \hspace{0.3cm} & $-0.0002$ \hspace{0.3cm}  \\ 
\hline
\end{tabular}
\caption{Atmospheric parameter values, and the temperature and density gradients are shown. Data are collected from Ref. \cite{usdata1}.}
 \label{table1} 
\end{table*}
%%%%%%%%%%%%%%%%%%%%%%%%%%
\par
\begin{figure*}
\centering
\includegraphics[width=6in, height=3in]{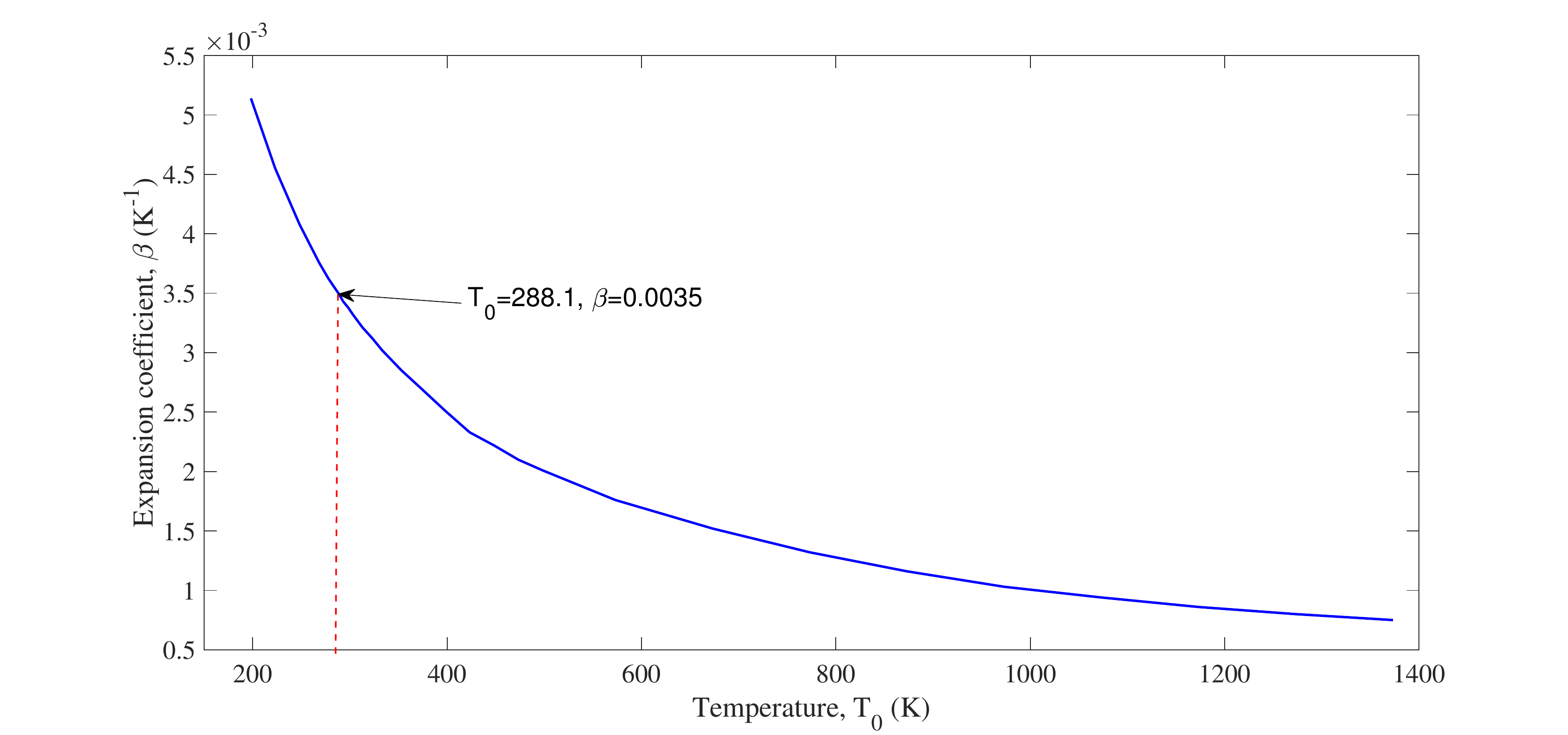}
\caption{The variation of the thermal expansion coefficient $(\beta)$ with the temperature $(T_0)$ is shown based on the data as in Ref. \cite{usdata2}. A critical value of $\beta$ at the maximum temperature $T_0\approx288.15$ K for which the instability occurs is indicated by the text arrow.  }
\label{fig2}
\end{figure*}

 
In what follows, we also show the dependence of the thermal expansion coefficient $(\beta)$ on the temperature $T_0$ in Fig. \ref{fig2}. The data used are as in Ref. \cite{usdata2}. It is clear that the expansion coefficient falls off quickly with increasing values of the temperature and that the maximum temperature $T_0\approx288.15$ K occurring at the Earth's surface corresponds to the thermal expansion coefficient $\beta\approx0.0035$. Later, we will show that such value of $\beta$ is minimum (corresponding to the maximum temperature $T_0\approx288.15$ K) above which the Brunt-V{\"a}is{\"a}l{\"a} frequency becomes negative $(N^2<0)$ and hence the instability of stratified fluid density perturbations. 
%%%%%%%%%%%% 
%%%%%%%%%
 \par 
 It is well known that the density variations due to internal gravity waves do not exceed $3-4\%$. So, the ratio between the density perturbation and the unperturbed density is small, i.e., $\rho_1/\rho_0\approx(1-4)\times10^{-2}$. In this case, the momentum equation \eqref{eq-mom} in the Boussinesq approximation reduces to
 \begin{equation}\label{eq-mom1}
 \frac{\partial {\bf u}}{\partial t}+\left({\bf u}\cdot\nabla\right){\bf u}=-\frac{1}{\rho_0}\nabla p_1-\frac{\rho_1}{\rho_0} g\hat{z}, 
\end{equation} 
where the suffix $1$ in $\rho$ and $p$ denotes perturbation and $\hat{z}$ is the unit vector along the $z$-axis. Next, using the relation \eqref{eq-rho},  Eq. \eqref{eq-mom1} reduces to \cite{tur2013a,tur2013b}
 \begin{equation}\label{eq-mom2}
 \frac{\partial {\bf u}}{\partial t}+\left({\bf u}\cdot\nabla\right){\bf u}=-\frac{1}{\rho_0}\nabla p_1+ g\beta T_1\hat{z}, 
\end{equation} 
where $T_1$ denotes the temperature perturbation.  Furthermore, we require the following heat equation for the imcompressible fluid in absence of any heat source \cite{tur2013a,tur2013b}.
\begin{equation}\label{eq-T}
 \frac{\partial T}{\partial t}+\left({\bf u}\cdot\nabla\right)T=\chi\nabla^2 T,
\end{equation}
where $\chi$ is the coefficient of the thermal diffusivity and is equal to the ratio between the therml conductivity $\kappa~\rm{(W/mK)}$ and the volumetric heat capacity $\rho C_p~\rm{(J/m^3K)}$. Here, $C_p$ is the specific heat capacity $\rm{(J/kg K)}$ and the mass density $\rho$ is in the unit of $\rm{kg/m^3}$. Representing the total temperature as the sum of its equilibrium and perturbed parts, i.e., $T=T_0(z)+T_1$, and assuming that $\alpha\equiv dT_0/dz$ as more or less a constant equilibrium gradient of temperature along the $z$-axis, i.e., $\nabla^2T=\nabla^2(T_0+T_1)=\nabla^2T_1$, from Eq. \eqref{eq-T} we obtain \cite{tur2013a,tur2013b}
\begin{equation}\label{eq-T1}
 \frac{\partial T_1}{\partial t}+\left({\bf u}\cdot\nabla\right)T_1=\chi\nabla^2 T_1-\alpha u_z,
\end{equation}
where $u_z$ is the component of ${\bf u}$ along the $z$-axis and $\alpha~(>0)$ represents the action of buoyancy force.   Equations \eqref{eq-mom1} and \eqref{eq-T1} with the  conditions  
\begin{equation} \label{eq-cont-compr1}
\nabla\cdot {\bf u}=0,~~\frac{d\rho}{dt}=0,
\end{equation}
are the desired set of equations for the evolution of the temperature and density perturbations of stratified incompressible fluids. 
\par
To elucidate the role of the temperature gradient (vertical), we consider the linear approximation, i.e.,  we consider the following simple model equations and remove the suffix $1$ in the perturbed variables, for simplicity. Separating the perpendicular and vertical (parallel to the gravity) components of Eq. \eqref{eq-mom2}, we obtain
\begin{equation}\label{eq-mom-perp}
\frac{\partial u_\perp}{\partial t}+\frac{1}{\rho_0}\nabla_\perp p=0,
\end{equation}
\begin{equation}\label{eq-mom-paral}
\frac{\partial u_z}{\partial t}+\frac{1}{\rho_0}\frac{\partial p}{\partial z}-g\beta T=0.
\end{equation}
Also, the equation $\nabla\cdot {\bf u}=0$ gives
\begin{equation} \label{eq-u}
\nabla_\perp\cdot {\bf u}=-\frac{\partial u_z}{\partial z}.
\end{equation}
Taking the gradient $(\nabla)$ of Eq. \eqref{eq-mom-perp}, noting that $\nabla_\perp^2=\Delta_\perp=\partial^2/\partial x^2+  \partial^2/\partial y^2$, and using Eq. \eqref{eq-u}, we get
\begin{equation} \label{eq-u1}
\frac{\partial^2u_z}{\partial t\partial z}=\frac{1}{\rho_0}\nabla_\perp p.
\end{equation}
Next, we operate $\partial \Delta_\perp/\partial t$ on Eq. \eqref{eq-mom-paral} to get
\begin{equation}\label{eq-u2}
\frac{\partial^2}{\partial t^2}\Delta_\perp u_z+\frac{1}{\rho_0}\frac{\partial^2}{\partial t\partial z}\nabla_\perp p-g\beta\frac{\partial}{\partial t}\Delta_\perp T=0.
\end{equation}
Furthermore, using Eq. \eqref{eq-u1} and noting that $\rho_0=\rho_0(z)$, from Eq. \eqref{eq-u2} we have 
\begin{equation} \label{eq-u3}
\frac{\partial^2}{\partial t^2}\left(\Delta u_z+\frac{1}{\rho_0}\frac{d\rho_0}{dz}\frac{\partial u_z}{\partial z}\right)-g\beta\frac{\partial}{\partial t}\Delta_\perp T=0,
\end{equation}
where the Laplacian operator, $\Delta=\Delta_\perp+\partial^2/\partial z^2$.
\par 
Also, operating Eq. \eqref{eq-T1} with $\Delta_\perp$, we get
\begin{equation}\label{eq-T3}
\frac{\partial}{\partial t}\Delta_\perp T=\chi\Delta_\perp\Delta T-\alpha\Delta_\perp u_z.
\end{equation}
Combining Eqs. \eqref{eq-u3} and \eqref{eq-T3} yields
 \begin{equation} \label{eq-u4}
\frac{\partial^2}{\partial t^2}\left(\Delta u_z+\frac{1}{\rho_0}\frac{d\rho_0}{dz}\frac{\partial u_z}{\partial z}\right)-g\beta\chi\Delta_\perp\Delta T+g\alpha\beta\Delta_\perp u_z=0.
\end{equation} 
Using the thermal expansion relation \eqref{eq-rho}, we recast the density conservation equation \eqref{eq-cont-compr1} as
\begin{equation}\label{eq-cont-compr2}
\left(1-\beta T_0-\beta T\right)\frac{d\rho_0}{d t}-\rho_0\beta\frac{d}{dt}\left(T_0+T\right)=0.
\end{equation}  
%%%%%%
By means of the heat equation \eqref{eq-T}, Eq. \eqref{eq-cont-compr2} gives, in the linear approximation, the following. 
\begin{equation}\label{eq-cont-compr3}
\left(1-\beta T_0\right)\frac{1}{\rho_0}\frac{d\rho_0}{d z}u_z =\beta\chi\Delta T.
\end{equation}
Finally, from Eqs. \eqref{eq-u4} and \eqref{eq-cont-compr3}, we obtain
\begin{equation}\label{eq-u5}
\frac{\partial^2}{\partial t^2}\left( \Delta u_z+\frac{1}{\rho_0}\frac{d\rho_0}{d z} \frac{\partial u_z}{\partial z}\right)+N^2\Delta_\perp u_z=0,
\end{equation}
where $N^2$ is the squared Brunt-V{\"a}is{\"a}l{\"a} frequency, given by,
\begin{equation}\label{eq-N2}
N^2(z)=g\left[\left(\beta T_0-1\right)\frac{1}{\rho_0}\frac{d\rho_0}{d z} +\beta \frac{dT_0}{d z} \right].
\end{equation}
%%%%%%%%%%%%%%%%%% 
Equation \eqref{eq-u5} represents a differential equation of only one unknown variable $u_z$ with the frequency $N^2$ being modified by the temperature stratification (proportional to $\beta$). In absence of the latter, one recovers the known Brunt-V{\"a}is{\"a}l{\"a} frequency \cite{acheson1973}.  Further simplification of Eq. \eqref{eq-u5} can be made by neglecting the second term in the parentheses,  compared to the first one. Thus, the dynamics of internal gravity waves in stratified fluids can be described by the following equation.
\begin{equation} \label{eq-u6}
\frac{\partial^2}{\partial t^2}\Delta u_z+N^2\Delta_\perp u_z=0.
\end{equation}
To elucidate the influence of the thermal expansion parameter $\beta$  on the stability of perturbations in vertical stratified fluids, from Eq. \eqref{eq-N2} we find that,  $N^2$  becomes negative when the thermal expansion coefficient $\beta$ satisfies the inequality:
\begin{equation}\label{eq-ineq0}
\beta T_0\left(L_{\rho_0}^{-1}+L_{T_0}^{-1} \right)<L_{\rho_0}^{-1},
\end{equation}
where $L_{\rho_0}^{-1}\equiv \left(1/\rho_0\right)|d\rho_0/dz|$ and $L_{T_0}^{-1}\equiv \left(1/T_0\right)|dT_0/dz|$, respectively,  denote the inverses of the length scales of density and temperature inhomogeneities.  Since in the altitudes of troposphere and stratosphere [$0<z<50$ (km)], $d\rho_0/dz<0$ (\textit{cf}. Table \ref{table1}), the inequality \eqref{eq-ineq0} reduces to 
\begin{equation} \label{eq-ineq1}
\beta T_0\left(\frac{1}{|L_{\rho_0}|}-\frac{1}{L_{T_0}} \right)> \frac{1}{|L_{\rho_0}|}.
\end{equation}
From Table \ref{table1}, it is also evident that $|L_{T_0}^{-1}|<|L_{\rho_0}^{-1}|$. Thus, from Eq. \eqref{eq-ineq1}, we get the following approximate condition of instability in vertical stratified fluids.
\begin{equation}\label{eq-ineq2}
\beta T_0>1.
\end{equation}
%%%%%%%%%%%%%%%%%%%%%%%%%%
From Table \ref{table1}, we find that the maximum value of the temperature is at the Earth's surface $(T_0\approx288.15$ K). So, the instability condition [Eq. \eqref{eq-ineq2}] holds for a minimum value of $\beta$: $\beta_{\rm{min}}\approx0.0035$. The latter well agrees with the observational data (See the text arrow in Fig. \ref{fig2}). 
\par
\begin{figure*}
\centering
\includegraphics[width=6in, height=3in]{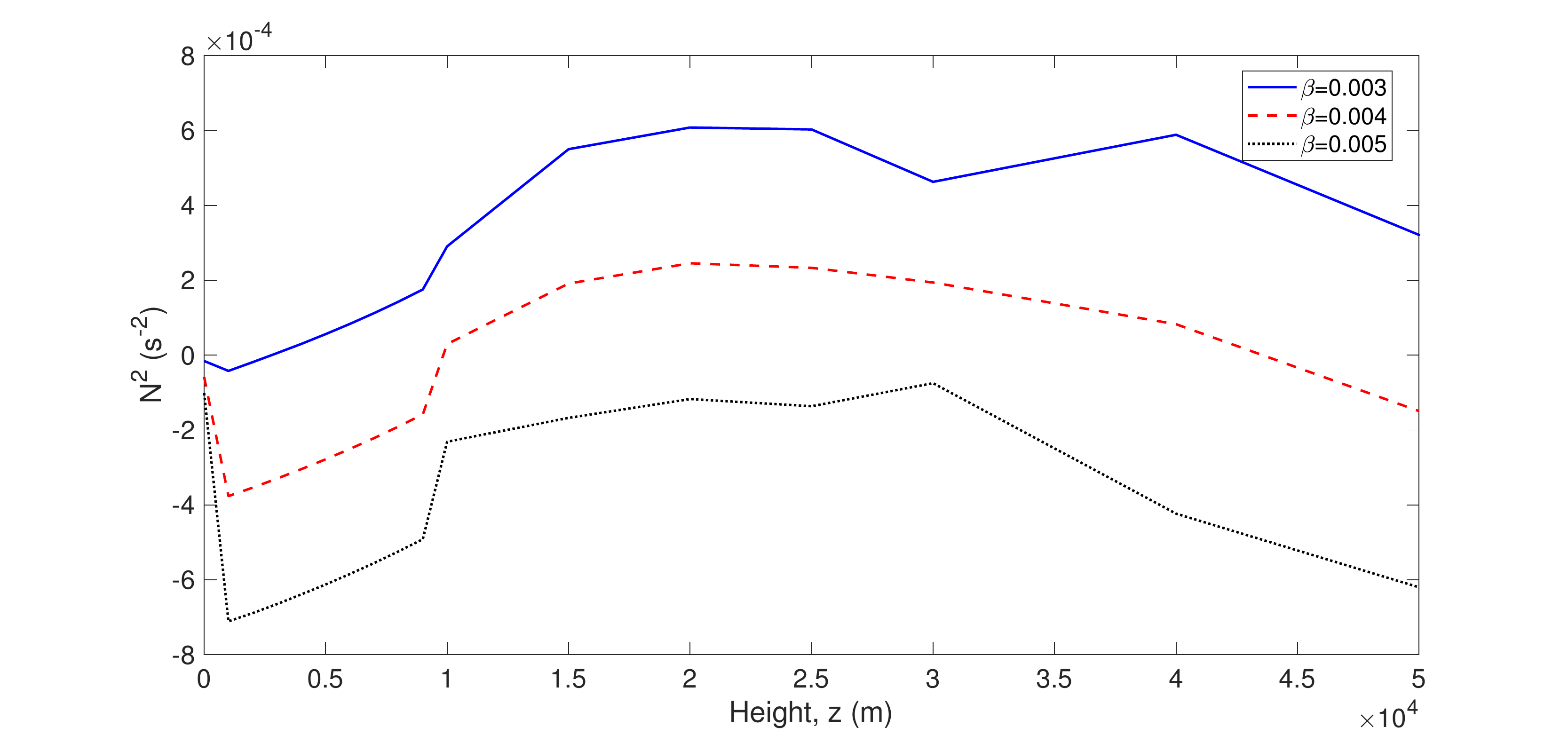}
\caption{The squared  Brunt-V{\"a}is{\"a}l{\"a} frequency $(N^2)$ is plotted against the height $(z)$ for different values of the thermal expansion coefficient $\beta$ as in the legend. The data used are as in Table \ref{table1}. It is seen that $N^2$ becomes negative and the instability of vertical stratification occurs for $\beta\gtrsim0.0035$.   }
\label{fig3}
\end{figure*}

 
 The dependence of the squared  Brunt-V{\"a}is{\"a}l{\"a} frequency $(N^2)$ on the thermal expansion coefficient $(\beta)$ is shown in Fig. \ref{fig3}. It is seen that the instability of  atmospheric stratification occurs with an increase of the thermal expansion coefficient beyond the critical value $(\approx0.0035)$.  The Brunt-V{\"a}is{\"a}l{\"a} frequency becomes completely negative for $\beta\gtrsim0.005$. In the latter, it is also noted that the magnitude of $N^2$ initially increases in the interval $0\lesssim z\lesssim10^3$ (m), and then decreases in $10^3\lesssim z\lesssim3\times10^4$ (m). In the rest of the interval, $3\times10^4\lesssim z\lesssim5\times10^4$ (m), its magnitude again increases. Such behaviors of $N^2$ may be due to the variation of the relative magnitudes of the length scales corresponding to the fluid density and temperature as the height $z$ increases from $z=0$ to $z=50$ km.  It is interesting to note that when the value of $\beta$ is lower than $\beta=0.005$, $N^2$ can be negative, zero, or positive depending on the altitude $z$. For example, when $\beta=0.003$, $N^2<0$ in $0\lesssim z\lesssim2\times10^3$ (m), $N^2\approx0$ at $z=3\times10^3$ (m), and $N^2>0$ in $3\times10^3\lesssim z\lesssim50\times10^3$ (m). Also, when $\beta=0.004$, $N^2<0$ in $0\lesssim z\lesssim9\times10^3$ (m), $N^2\approx0$ at $z=10\times10^3$ (m),  and $N^2>0$ in $15\times10^3\lesssim z\lesssim30\times10^3$ (m). Again, $N^2\approx0$ at $z=40\times10^3$ (m), and $N^2<0$ at $z=50\times10^3$ (m).   Physically, when $N^2>0$, Eq. \eqref{eq-u6} admits oscillating solutions for the velocity $u_z$ with frequency $N$, i.e., if a parcel of stratified neutral fluids moves upward and $N^2>0$,  it will oscillate in between the heights where the fluid density of the parcel matches with the surrounding fluids. In this case, the fluid is said to be stable. However, when $N^2=0$, the parcel, once pushed up, will not move any  further. On the other hand, when $N^2<0$, i.e., the squared  Brunt-V{\"a}is{\"a}l{\"a} frequency becomes imaginary, the parcel will move up and up   until $N^2$ becomes zero or positive again in the atmosphere. Typically, such a situation leads to convection, and hence the criterion for the stability of stratified fluids in the atmosphere against convection is that $N^2>0$.   
%%%%%%%%%%%%%%%%%%%%%%%%%%%%%%%%%%%%%%

%%%%%%%%%
%%%%%%%
\section{Conclusion}
We have studied the influence of the thermal expansion of the Earth's atmosphere on the stability of vertical stratification of density and temperature perturbations. We have shown that such an influence can lead to instability in stratified incompressible fluids. Modified by the thermal expansion coefficient, the Brunt-V{\"a}is{\"a}l{\"a} frequency is obtained, and a critical value of the expansion coefficient for which the instability occurs is revealed.  
\par To conclude, the instability of vertical stratification reported here could be helpful for the  initiation of large-scale instability (which may be larger than the scales of any external force or turbulence phenomena) as well as the generation of large-scale vortices in the atmosphere \cite{kopp2021} through which the particle momentum and energy transfer take place. In the fluid model, we have neglected the dissipative effects, such as those associated with the fluid-particle collision and the kinematic viscosity. These effects will contribute to the evolution equation for internal gravity waves, modify their dispersion properties, and may eventually reduce or prevent the instability of stratified fluids reported here. However, the influence of these forces and the effects of the temperature and density gradients on the propagation characteristics of internal gravity waves are beyond the scope of the present work but a project for our future study.

 \section*{Declaration of Competing Interest}
The authors declare that they have no known competing financial interests or personal relationships that could have appeared to influence the work reported in this paper.
%\section*{Acknowledgments}

% \bibliographystyle{elsarticle-num} 
 %\bibliographystyle{cas-model2-names}
\bibliographystyle{apsrev4-1}
\bibliography{Reference}

\end{document}